\documentclass[twocolumn,showpacs]{revtex4}
\usepackage{graphicx}
\usepackage{dcolumn}

\begin{document}

%\preprint{APS/123-QED}

\title{Gapless quantum spin liquid, stripe and antiferromagnetic phases in frustrated Hubbard models in two dimensions}

\author{Takahiro Mizusaki}
\affiliation{
Institute of Natural Sciences, Senshu University, 
Kanda, Chiyoda, Tokyo 101-8425, Japan
}%Lines break automatically or can be forced with \\

\author{Masatoshi Imada$^{1,2}$}%
\affiliation{%
${^1}$Institute for Solid State Physics, University of Tokyo,
Kashiwanoha, Kashiwa, 277-8581, Japan and Department of Applied Physics, 
University of Tokyo,
Hongo, Bunkyo-ku, Tokyo, 113-8656, Japan \\
${^2}$PRESTO, Japan Science and Technology Agency}%%

\begin{abstract}
Unique features of non-magnetic insulator phase are revealed, and
the phase diagram of the $t-t'$ Hubbard model containing the diagonal transfers $t'$ on a square lattice is 
presented. 
Using the path-integral renormalization group method, we find an
antiferromagnetic phase for small next-nearest neighbor transfer $t'$ and a
stripe (or collinear) phase for large $t'$ in the Mott insulating region of the
strong onsite interaction $U$. For intermediate $t'/t\sim 0.7$ at large 
$U/t>7$, we find a longer-period antiferromagnetic-insulator phase with 
$2\times 4$ structure. In the Mott insulating region, we also find a quantum 
spin liquid (in other words, a non-magnetic insulator) phase near the Mott 
transition to paramagnetic metals for the $t-t'$ Hubbard model on the square lattice as well as on the anisotropic triangular lattice.
Correlated electrons often crystallize to the Mott insulator usually with some magnetic orders, whereas the ``quantum spin liquid" has been a long-sought issue.  
We report numerical evidences that a nonmagnetic insulating (NMI) phase gets stabilized near the Mott transition with remarkable properties:
 The 2D Mott insulators on geometrically frustrated lattices contain a phase with gapless spin excitations and degeneracy of the ground state in the whole Brillouin zone of the total momentum. The obtained vanishing spin renormalization factor suggests that spin excitations do not propagate coherently in contrast to the conventional phases, where there exist either magnons in symmetry broken phases or particle-hole excitations in paramagnetic metals.  It imposes a constraint on the possible pictures of quantum spin liquids and supports an interpretation for the existence of an unconventional quantum liquid.
The present concept is useful in analyzing a variety of experimental results in frustrated magnets including organic BEDT-TTF compounds and $^3$He atoms adsorbed on graphite.
\end{abstract}


\maketitle

%=================================================================
\section{Introduction}\label{sec1}
Among various insulating states, those caused by electronic Coulomb correlation effects, called 
the Mott insulator, show many remarkable phenomena such as high-$T_c$ superconductivity and colossal magnetoresistance near it~\cite{Mott,RMP}. However, it has also been an issue of long debate whether the Mott insulator has its own identity distinguished from insulators like the band insulator.
This is because the Mott insulator in most cases shows symmetry breakings such as antiferromagnetic order or dimerization, where the resultant folding of the Brillouin zone makes the band full and such insulators difficult to distinguish from the band insulators
because of the adiabatic continuity.

Except for one-dimensional systems, the possibility of the inherent Mott insulator without conventional orders has been a long-sought challenge.  
The Mott insulator on the triangular lattice represented by the Heisenberg spin systems was proposed as a candidate~\cite{Anderson}. 
Although, the triangular Heisenberg system itself has been argued to show an antiferromagnetic (AF) order~\cite{Lhuillier}, intensive studies on geometrical frustration effects have been stimulated.  
In particular, the existence of a quantum spin liquid phase has been 
established in recent unbiased numerical studies performed on two-dimensional
lattices with geometrical frustration effects\cite{Kashima,Morita,IMW}.
The spin liquid has been interpreted to be stabilized by charge fluctuations enhanced near the Mott transition.

Recently extensive experimental studies on frustrated quantum magnets such as those on triangular, Kagome, spinel and pyrochlore lattices~\cite{Ramirez,Greedan,Kanoda,Kato,Hiroi,Taguchi,Wiebe} as well as on triangular structure of $^3$He on graphite~\cite{Ishida} have been performed. 
They tend to show 
suppressions of magnetic orderings with large residual entropy with
a gapless liquid feature for quasi 2D systems or ``spin glass-like" behavior in 3D even for disorder-free compounds. These gapless and degenerate behaviors wait for a consistent theoretical understanding.

In this paper, by extending and reexamining the previous studies~\cite{Kashima,Morita,IMW}, we further show more detailed numerical evidence for the existence of a new type of inherent Mott insulator {\it near the Mott transition}; singlet ground state with unusual degeneracy, namely gapless and dispersionless spin excitations.   
Our present results offer a useful underlying concept for the understanding of the puzzling feature in the experiments.   

In this paper, we also show that several different antiferromagnetic phases appear in the region of large $U/t$.  This includes normal antiferromagnetic order stabilized for small $t'/t$ with the Bragg wavenumber $Q=(\pi,\pi)$, collinear (stripe) order for large $t'/t$ with $Q=(0,\pi)$ and longer-period antiferromagnetic order for intermediate $t'/t$ with $Q=(\pi,\pi/2)$. 

\section{Frustrated Hubbard Models}

\begin{figure}[h]
\includegraphics[width=4.0cm]{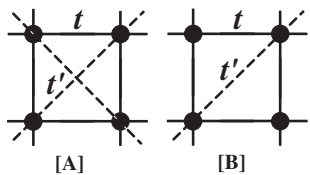}
\caption{Lattice structure of geometrically frustrated lattices on a square lattice.  The nearest- and next-nearest-neighbor transfers are denoted by $t$ and $t'$, respectively.}
\label{Fig1}
\end{figure}

In the present study, we investigate the Hubbard model on two-dimensional frustrated lattices.  
The Hamiltonian by the standard notation reads 
\begin{eqnarray}  
H&=&H_K+H_U
\nonumber\\
H_K&=&-\sum_{\langle i,j \rangle ,\sigma}t   
\left(c_{i\sigma}^{\dagger}c_{j\sigma}+H.c.\right)
  + \sum_{\langle k,l \rangle,\sigma}t' \left(c_{k\sigma}^{\dagger}c_{l\sigma}+{\rm H.c.}\right)
\nonumber\\ 
H_U&=&U\sum_{i=1}^{N} \left(n_{i\uparrow}-\frac{1}{2}\right)  
\left(n_{i\downarrow}-\frac{1}{2}\right) 
\label{Hamiltonian}  
\end{eqnarray} 
on a $N$-site square lattice with a nearest neighbor ($t$) and two choices of diagonal next-nearest neighbor ($t'$) transfer integrals in the configuration illustrated in Fig.~\ref{Fig1}[A] and [B]. The energy unit is taken by $t$. Hereafter we call the Hubbard models with the lattice structures illustrated in Fig.~\ref{Fig1}[A] and [B], the models [A] and [B], respectively. 
The $t'$ transfer integrals bring about geometrical frustration.
The $i$, $j$ represent lattice points 
and  $c_{i\sigma }^{\dagger}$ ($c_{j\sigma }$) 
is a creation (annihilation) operator of an electron with spin $\sigma$
on the $i$-th site.

%===============  fig. 1  ========================================

%=================================================================

\section{Path-Integral Renormalization Group method}

The model requires accurate and unbiased theoretical calculations because of large fluctuation effects expected from the low dimensionality of space and the geometrical frustration effects due to nonzero $t'$.    
Recently, the path-integral renormalization group (PIRG) method~\cite{Kashima-Imada} opened a way of numerically studying the models with the frustration effects more thoroughly without the negative sign problem and without relying on the Monte Carlo sampling. The efficiency of the method was established through a number of applications~\cite{Kashima,Morita,Noda,MizusakiN}. 
Furthermore, a quantum number projection method has been introduced into
the PIRG method and symmetries of the model can be handled explicitly
and precision of solutions can be significantly enhanced \cite{QPPIRG}.

Here we briefly introduce the PIRG method and quantum number projections.
The ground state $| {\psi _g} \rangle$ can be, in general, 
obtained  by applying the projector
$e^{-\tau H }$ to an arbitrary state $| {\phi _{\rm initial}} \rangle $ 
which is not orthogonal to the true ground state as
\begin{equation}
| {\psi _g} \rangle=e^{-\tau H }| {\phi _{\rm initial}} \rangle. 
\end{equation}
To operate $\exp[-\tau H]$, we decompose $\exp[-\tau H]$ into $\exp[-\tau H] \sim [\exp[-\Delta\tau H_K]\prod_i \exp[-\Delta\tau H_{U_i}]]^{\cal N}$ for small $\Delta\tau$, where $\tau={\cal N}\Delta\tau$.
When we use the Slater determinant as the basis functions, the operation of $\exp[-\Delta\tau H_K]$ to a Slater determinant simply transforms to another single Slater determinant. On the other hand, the operation of $\exp[-\Delta\tau H_{U_i}]$ can be performed by the Stratonovich-Hubbard transformation, where a single Slater determinant is transformed to a linear combination of two Slater determinants. Therefore, the operation of  $\exp[-\tau H]$
increases the number of Slater determinants.
To keep manageable number of Slater determinants in actual computation, 
we restrict number of Slater determinants by selecting variationally
better ones. This process follows an idea of the renormalization group in the wavefunction form. 
Its detailed algorithm and procedure are found in Ref.~\cite{Kashima-Imada}.
After the operation of $\exp[-\tau H]$, the projected wave function can be given by an optimal form
composed of $L$ Slater determinants as
\begin{equation}
| {\psi ^{(L)}} \rangle =\sum_{\alpha =1}^L {c_\alpha }| 
{\phi^{(L)}_\alpha } \rangle, 
\label{multi-basis}
\end{equation}
where $c_\alpha$'s are amplitudes of $| {\phi^{(L)}_\alpha } \rangle $.
Operation of the ground-state projection  can give optimal $c_\alpha$'s 
and $| {\phi^{(L)}_\alpha } \rangle $'s for a given $L$.

In most cases, eq. (3) can give only upper-bound of the exact energy 
eigenvalue.
Therefore, to obtain an exact energy, we consider
an extrapolation method based on a relation between energy 
difference $\delta E$ and energy variance $\Delta E$
\cite{Kashima-Imada,sorella}.
Here the energy difference is defined as
$
\delta E = \langle {\hat H} \rangle - \langle {\hat H} \rangle_g
$
and the energy variance is defined as
$
\Delta E={\frac{\left\langle {\hat H^2} \right\rangle -\left\langle {\hat H} \right\rangle ^2}{\left\langle {\hat H} \right\rangle ^2}}.
$
Here, $\langle {\hat H} \rangle_g$ stands for the true ground-state energy.
For $\left| {\psi ^{(L)}} \right\rangle$, we evaluate the energy $E^{(L)}$ and energy  variance $\Delta E^{(L)}$, respectively.

If $\left| {\psi ^{(L)}} \right\rangle$ is a good approximation of the true state,
the energy difference $\delta E^{(L)}$ is proportional to the energy 
variance $\Delta E^{(L)}$.
Therefore extrapolating $E^{(L)}$ into $\Delta E^{(L)}\to 0$ 
by increasing $L$ systematically, we can accurately estimate the 
ground-state energy.

Next we consider a simple combination of the PIRG and quantum number projection. 
The PIRG gives approximate wave function for a given $L$ which is 
composed of $L$ linear combinations of  
$\left| {\phi^{(L)}_\alpha } \right\rangle $.
%This wave function still often breaks symmetries due to limitation of numerically manageable number $L$.  
Though spontaneous symmetry breaking should not occur
in finite size systems, symmetries are sometimes broken in PIRG calculations
because of the limited number of basis functions, if symmetry projections
are incompletely performed. However, extrapolations to the
thermodynamic limit recover the true ground state, in which possible 
symmetry breakings are correctly evaluated.

For finite size system, to handle wave functions with definite and exact symmetries,
we can apply quantum number projection to this wave functions as 
\begin{equation}
{\bf P}\left| {\psi ^{(L)}} \right\rangle =
\sum\limits_{\alpha =1}^L {c_\alpha }{\bf P}\left| {\phi^{(L)}_\alpha } \right\rangle, 
\end{equation}
where $\bf P$ is a quantum-number projection operator.
We can use the same amplitudes $c_\alpha$'s and the same bases  
$\left| {\phi^{(L)} _\alpha } \right\rangle$'s which the PIRG determines, 
while this amplitude $c_\alpha$'s can be easily reevaluated by
diagonalization by using quantum-number projected bases, that is, 
we determine $c_\alpha$'s by solving the generalized eigenvalue 
problem as
\begin{equation}
H^{\bf P}_{\alpha \beta }\vec x = N^{\bf P}_{\alpha \beta }\vec x, 
\end{equation}
where 
$N^{\bf P}_{\alpha \beta }=\left\langle {\phi _\beta } \right|{\bf P}\left| 
{\phi _\alpha } \right\rangle $, 
$H^{\bf P}_{\alpha \beta }=\left\langle {\phi _\beta } \right|H{\bf P}\left| 
{\phi _\alpha } \right\rangle $.
The latter procedure gives a lower energy eigenvalue.
By adding this procedure for the PIRG basis, we evaluate the projected energies and
energy variances, $E^{L}_{\rm proj}$ and $\Delta E^{L}_{\rm proj}$ for each $L$. We can estimate
accurate energy by extrapolating the projected energy into zero variance.
Consequently we can exactly treat the symmetry and extract the state with specified quantum numbers by the PIRG.
We call this procedure PIRG+QP.

%=================================================================

\section{Phase diagram on $t'-U$ plane}

For $t'=0$, the metal-insulator transition occurs at $U=0$. The $t'$ offers additional 
dimension for parameter spaces of Hubbard models. Recently the existence of a nonmagnetic insulator (NMI) near the metal-insulator transition boundaries was reported~\cite{Kashima,Morita} on the two-dimensional frustrated Hubbard model on a lattice by using the PIRG method.
In the present paper, we thoroughly investigate two-dimensional parameter spaces
spanned by $U/t$ and $t'/t$. 

\begin{figure}[h]
\includegraphics[width=9.0cm]{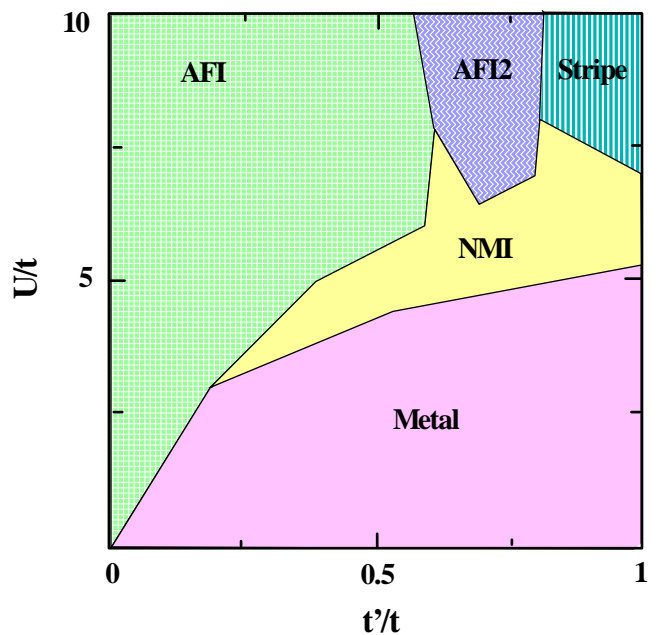}
\caption{(color online) Phase diagram of the Hubbard model with the lattice structure illustrated in Fig. 1 [A]
in the parameter space of $U$ scaled by $t$, and the frustration parameter $t'/t$.  AFI, AFI2, Stripe, PM, and NMI represent two types of antiferromagnetic insulating, 
stripe shape insulating,
paramagnetic metallic and nonmagnetic insulating phases, respectively.}
\label{Fig2}
\end{figure}

To obtain the ground state for each point ( $U/t$, $t'/t$ ), we carry out 
calculations for $6 \times 6$, $8 \times 8$ and $10 \times 10$ lattices
by the PIRG+QP method.
By extrapolating the ground state energies per lattice site, we determine its thermodynamic value. 
By a thorough search of the parameter space and improved accuracy of the computation, the phase diagram is clarified for the model [A] in a wider region of the parameter space than the previous study~\cite{Kashima}. It becomes quantitatively more accurate and contains a new feature, which has not been revealed in the previous results~\cite{Kashima}.  
In Fig.~\ref{Fig2}, we show the phase diagram, where various types of phase appear. 
For small $U/t$, paramagnetic metallic phase appears and for larger $U/t$ and 
small $t'/t$,  antiferromagnetic insulating phase appears. 
As $t'/t$ increases, non-magnetic insulating phase appears.
This feature has already been reported in Ref.\cite{Kashima}.
In addition to these phases, we find two additional phases: one is another type of 
antiferromagnetic insulating phase and the other is stripe ordered insulating phase.
The nature of these new phases will be reported in the following sections. 

We determine the phase boundary in the following way.
The ground state energies per site in the thermodynamic limit are obtained 
for each mesh point of this parameter space at 
$
\left( {U/t,t'/t} \right)=
\left( {n_1,n_2 \times 0.1} \right)
$
where $n_1,n_2=1,...10$ after the size extrapolation.
The phase boundary is determined by interpolating these energies per site.
 
\section{AFI phase}

\begin{figure}[h]
\includegraphics[width=8.0cm]{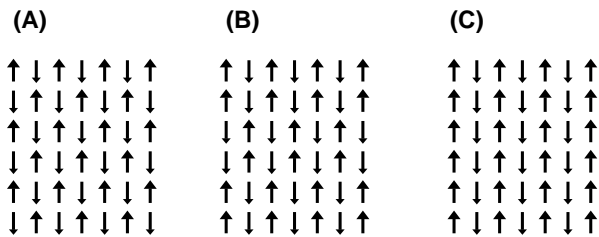}
  \caption{Configurations of AFI phase (A), new AFI phase (B) and Stripe phase (C).}
\label{config}
\end{figure}

For $t'/t=0$, the AFI phase of the configuration illustrated in Fig.~\ref{config} (A)  appears.
By the PIRG+QP method, we can evaluate an energy gap between the ground state ($S=0$) and the first excited state ($S=1$) for AFI phase.

\begin{figure}[h]
\includegraphics[width=8.3cm]{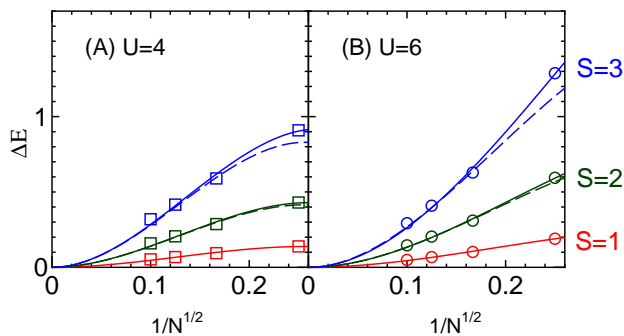}
\caption{(color online) Size scalings of the energy gaps for total spin $S$=1,2 and 3 in the AFI phase (A) ($U=4,t=1,t'=0$) and (B) ($U=6,t=1,t'=0$).  The solid curves are fitting by the form (6) and the dashed curves illustrate curves obtained from the $S=0$ fitting multiplied with the factor $4S(S+1)/3$.}
\label{Fig3}
\end{figure}

The finite-size gap for an $\ell \times \ell$ system in the chiral perturbation theory~\cite{Hasenfratz-Niedermayer} in the form
\begin{eqnarray}  
\Delta E= \frac{c^2}{\rho \ell^2}[1-\frac{3.900265c}{4\pi\rho \ell}+O(\frac{1}{\ell^2})]
\label{Spinwave}    
\end{eqnarray} 
is fitted with the calculated results in Fig.~\ref{Fig3}. The finite-size gap nicely follows the form (\ref{Spinwave}) in the AFI phase. For example, at $U=4, t=1$ and $t'=0$, the fitting in Fig.~\ref{Fig3} shows the spin wave velocity $c\sim 0.74$ and the spin stiffness $\rho \sim c/8.15$, which are equivalent to the estimate of the Heisenberg model at the exchange coupling $J=0.45$ in the spin wave theory. 
The fitted values of $c$ and $J$ well reproduce the previous estimates (ex. $J \sim 0.4$) obtained from the susceptibility and the staggered magnetizations~\cite{Hirsch}.
For $U/t=6.0$, we obtain the spin wave velocity $c\sim 0.97$ and the spin stiffness $\rho \sim c/5.44$, while for $U/t=8.0$, $c\sim 1.03$ and the spin stiffness $\rho \sim c/4.61$.
The excitations in the AFI phase well satisfy the tower structure of the low-energy excitation spectra based on the nonlinear sigma model description. 
The AFI phase is extended in the region of nonzero and moderate amplitude of $t'/t$ with large $U/t$ in the phase diagram.

\section{AFI phase with longer period}

In Fig.2, AFI phase with longer period (AFI2) appears. 
Its schematic configuration is depicted 
in Fig.~\ref{config}(B).

\begin{figure}[h]
\includegraphics[width=8.0cm]{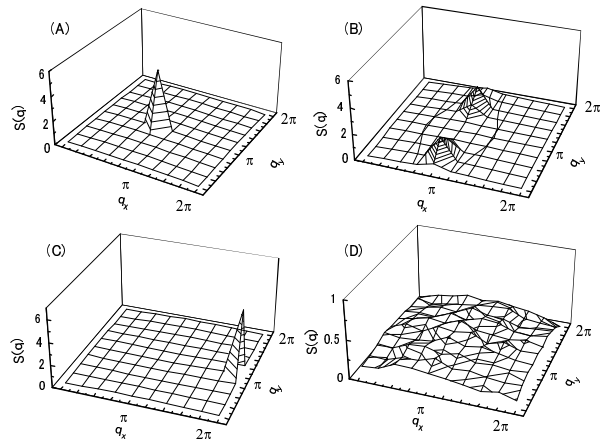}
  \caption{Equal-time spin correlations in the momentum space on 
  $10 \times 10$ at half filling for AFI phase (A) $(U=6.0,t'/t=0.5)$, new AFI phase (B) $(U/t=9.0,t'/t=0.7)$, stripe phase (C) $(U/t=9.0,t'/t=1.0)$ and NMI phase (D) $(U/t=6.0,t'/t=0.7)$.
  }
  \label{sdwpeaks}
\end{figure}

\begin{figure}[h]
\includegraphics[width=8.0cm,height=7cm]{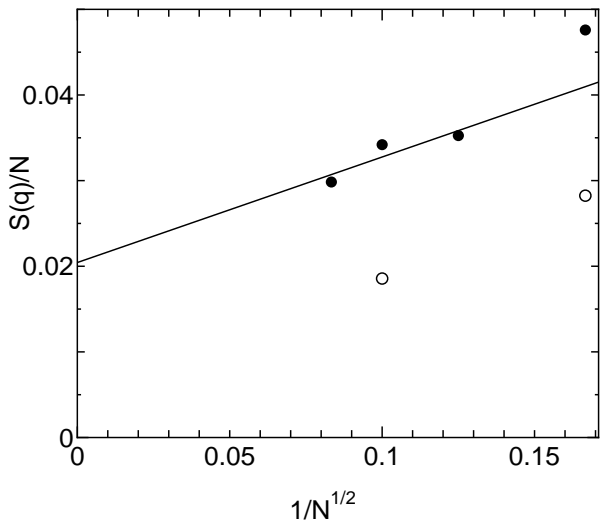}
\caption{
Finite-size scaling of $S(\pi,\pi/2)$ for 
$U/t=9.0$ and $t'/t=0.7$ at half filling.  The result suggests $2\times 4$ AF long-range order. 
Open circles for $6\times 6$ and $10\times 10$ 
lattices
are the peak values while filled symbols are the intensity of peak summed 
over
$k$ points at $(\pi,\pi/2)$ and its nearest neighbor $k$ points.
}
\label{sdw-nafi}
\end{figure}

To examine this configuration, we consider the 
equal-time spin structure factor defined, in the momentum space, by 
\begin{equation}
S\left( q \right) = \frac{1}{{3N}}\sum\limits_{i,j}^N {\left\langle {{\bf{S}}_i  \cdot {\bf{S}}_j } \right\rangle } e^{iq\left( {R_i  - R_j } \right)} 
\end{equation}
where $S_i$ is the spin of the $i$-th site and $R_i$  is the vector 
representing the coordinate of the $i$-th site.
In Fig.~\ref{sdwpeaks}(B), momentum dependence of $S(q)$ for the ground state at $U/t=9.0$ and $t'/t=0.7$ is shown, where we see distinct peaks 
at $(\pi,\pi/2)$ and $(\pi,3\pi/2)$.
In Fig. \ref{sdw-nafi}, we plot finite-size scaling of these peak amplitudes, which indicates the existence of the long-ranged order.
A usual AFI phase has a configuration pattern in Fig. \ref{config}(A),
which has $2 \times 2$ super structure. 
On the other hand, this new AFI phase has $2 \times 4$ super structure. 
Therefore, for smaller lattices, it is difficult to
identify it. For $4n \times 4n$ ($n$ is an integer) lattice, this structure is 
naturally realized and peak positions of spin correlation sharply appear at $(\pi,\pi/2)$ and $(\pi,3\pi/2)$,
while for $(4n+2) \times (4n+2)$ ($n$ is an integer) lattice,
peak positions of spin correlation  become wider 
from $(\pi,\pi/2-\delta)$ to $(\pi,\pi/2+\delta)$
and from $(\pi,3\pi/2-\delta)$ to $(\pi,3\pi/2+\delta)$
($\delta$ is around $\pi/4$). 
Thus there is an irregularity in spin correlation as a function of lattice size.
In Fig.~\ref{sdw-nafi}, we show summed amplitude of $S(q)$ over the peak for $6 \times 6$ and $10 \times 10$ lattices.


For each basis $| {\phi} \rangle $
of Eq.~(\ref{multi-basis}), we apply shift operation $T_{\Delta i,\Delta j}$ on the lattice and evaluate an overlap 
$\left\langle \phi  \right|T_{\Delta i,\Delta j} \left| \phi  \right\rangle $.
This overlap becomes quite large when $\Delta i=2$ or $\Delta j=4$, which reflects
a basic configuration in Fig. \ref{config} B.

In the strong-correlation limit ($U/t\rightarrow \infty$), 
the frustrated Hubbard models become the $J_1$-$J_2$ Heisenberg model 
in the leading order by
\begin{equation}
 H = J_1\Sigma_{\langle i,j\rangle} S_i\cdot S_j +
 J_2\Sigma_{\langle\langle i,j\rangle\rangle} S_i\cdot S_j 
\end{equation}
where $\langle \rangle$ and $\langle\langle\rangle\rangle$ denote the nearest-neighbor sites 
and next-nearest-neighbor sites, respectively. 
Here, $J_1=4t^2/U$ and $J_2=4t'^2/U$ are both antiferromagnetic 
interactions so that magnetic frustration arises. 
For $J_2/J_1<0.4$ the Neel order with peak structure in 
$S(q)$ at $q=(\pi, \pi)$ was proposed, which corresponds to
the AFI phase in the Hubbard model. 
On the other hand, for $J_2/J_1>0.6$,  the stripe order 
with $q = (0, \pi)$ or $(\pi, 0)$ peak in $S(q)$ was proposed, which corresponds to
the stripe shape \cite{Stripe} in the next section.
For the intermediate region of $0.4<J_2/J_1<0.6$, 
no definite conclusion has, however, been drawn on the nature of the ground state.
The possibility of the columnar-dimerized state
\cite{columnar1,columnar2,columnar3}, 
the plaquette singlet state \cite{Plaquette}
and the resonating-valence-bond state were discussed. 
In the present study, we find new type of AFI phase with longer period between AFI and stripe phases.
This new AFI phase could have some connection to the region $0.4 < J_2/J_1 < 0.6$.
In terms of the effects of frustrations, the stabilization of the longer-period AF structure is a natural consequence in the classical picture, where well known ANNNI model exhibits a devil's staircase structure\cite{Fisher}. Complicated longer-period structure may melt when quantum fluctuations are switched on, while  the present result indicates the survival of $2\times 4$ structure for large $U$.

\begin{figure}[h]
\includegraphics[width=8.0cm]{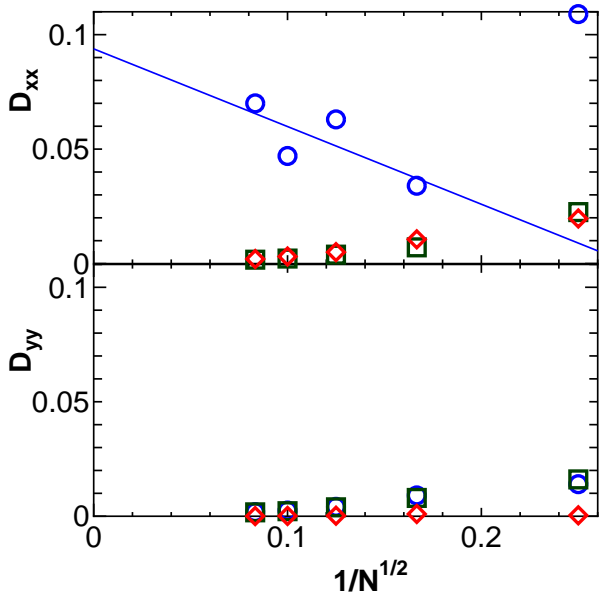}
  \caption{(color online) System-size dependence of the dimer-correlation
functions ($D_{xx}$ and $D_{yy}$) for $t$ = 1.0, $t'$= 0.7 and $U = 9.0$ at half filling on
$N = 4 \times 4, 6 \times 6, 8 \times 8, 10 \times 10$ and $12 \times 12$
lattices. 
The blue open circles, the red open diamonds and green open squares
show the $D_{xx}$ and $D_{yy}$ of AF2, Stripe and AF phases, 
respectively.
The long-range order suggested in $D_{yy}$ implies the $2\times 4$ 
AF order.}
\label{dimer}
\end{figure}

To investigate the possibility of the dimerized state
and the plaquette state, we consider the dimer correlation function 
$D_{\alpha ,\beta } $ for $\alpha, \beta=x,y$ 
defined by 
\begin{equation}
D_{\alpha ,\beta }  = \left\langle {O_\alpha  O_\beta  } \right\rangle 
\end{equation}
where
\begin{equation}
O_\alpha   = \frac{1}{N}\sum\limits_{i = 1}^N {\left( { - 1} \right)^i S_i  \cdot S_{i + \hat \alpha } } 
\end{equation}
and $\hat{\alpha}$ shows the unit vector in the $\alpha$ direction.
$D_{yy}$ clearly indicates that the dimer order in the $y$ direction is absent as we see in Fig.~\ref{dimer}. On the other hand, in this definition of $D$, $D_{xx}$ should also show the long-range order, if the $2\times 4$ AF order is stabilized.  This is indeed seen in our data. 
In the limit of strong coupling, this region is mapped to the Heisenberg 
model
with nearest-neighbor exchange $J_1$ and the next-nearest-neighbor exchange 
$J_2$,
with $J_2/J_1\sim 0.5$.  In this region, columnar dimer state has been 
proposed as the
candidate of the ground state~\cite{columnar1,columnar2,columnar3}.
Our result also shows the dimer long-range order supporting the existence of 
the columnar
dimer phase. However, in this phase at finite $U$, the antiferromagnetic 
long-range order with longer
period with $2\times 4$ structure shown in Fig. \ref{config} (B) is also 
seen. Since the dimer-order parameter defined by $D_{xx}$ is automatically 
nonzero in this extended antiferromagnetic phase, the primary order 
parameter should be identified as the longer-period antiferromagnetic order. 
We naturally expect a continuous connection of the Hubbard model to the 
Heisenberg model, while as far as we know, serious examination of such 
longer period AF order is not found in the literature.  It is an intriguing 
issue to examine this new possibility of longer-period AF order in the 
Heisenberg limit.
%The result indicates that $2\times 4$ AF order is realized rather than the dimer order.

\section{Stripe phase}
\begin{figure}[h]
\includegraphics[width=8.0cm]{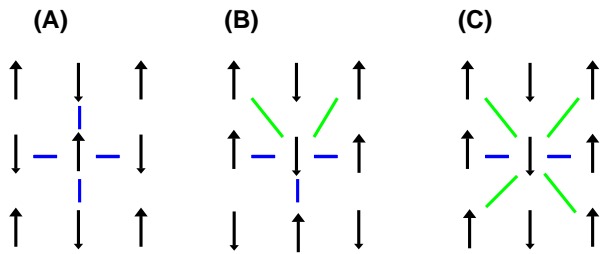}
  \caption{(color online) Strong AF bonds of AFI phase (A), new AFI phase (B) and 
  stripe phase (C).}
  \label{bond}
\end{figure}

For larger $t'/t$ and $U/t$, another phase appears, whose basic configuration
is stripe shape in Fig.~\ref{config}(C).
In Fig. \ref{bond}, we show strong AF bonds for $t$ and $t'$ directions.
For AFI phase, spins on the bonds along $t$-direction become antiferromagnetic each other and gain energy while spins on the bonds in $t'$-direction become parallel and lose energy.  On the other hand, for the stripe shape,
spins on the bonds for $t'$-direction become
all anti-parallel with compromised anti-parallel spins on the bonds for $t$-direction. 
%Therefore, larger $U/t$ means fewer double occupation on single site. 
In this situation, as $t'/t$ becomes larger,
stripe phase becomes energetically favored. 
As shape (B) is between (A) and (C), it becomes energetically favored
at medium value of $t'/t$. 

\begin{figure}[h]
\includegraphics[width=8.0cm]{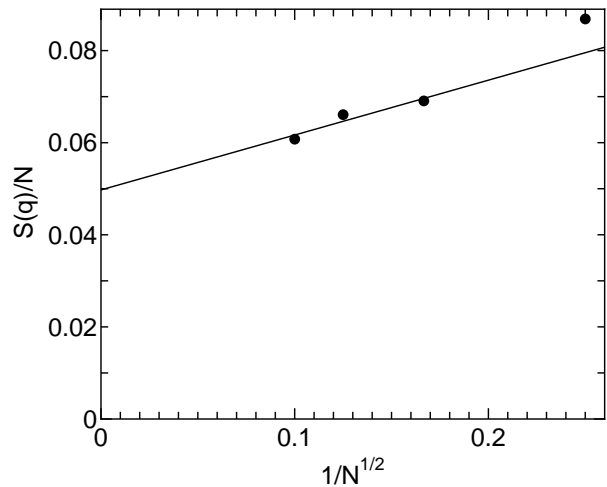}
\caption{Finite-size scaling of $S(0,\pi)$ 
for $U/t=9.0$ and $t'/t=1.0$ at half filling. The result suggests that the stripe has long-range order at this parameter value.}
\label{sdw-stripe}
\end{figure}

The minimum block of stripe shape is $2 \times 1$. Therefore,
there is no irregularity of spin correlation for different lattice sizes.
The spin correlation has a sharp peak at $(0,\pi)$ as we see in Fig.~\ref{sdwpeaks}(C).
Finite-size scaling in Fig.\ref{sdw-stripe} indeed shows the presence of the long-range order in the thermodynamic limit.
The overlap between basis state and its shifted one also indicates
that the configuration in Fig. \ref{config}(C) is realized.  
 
\section{Non-magnetic insulator phase}

Three phases of previous sections are semi-classical.
Their main configurations can be intuitively understood from the classical picture of the spin alignment.
Next we consider an unconventional phase of frustrated Hubbard models, which is illustrated in Fig.\ref{Fig2} as the nonmagnetic insulator (NMI). This phase does not have any distinct peak structure in $S({\bf k})$ as we see in Fig.~\ref{sdwpeaks}(D).
Because this NMI phase appears in a window sandwiched by the metal and 
insulating antiferromagnetic phases,
it is clearer than the previous studies\cite{Kashima,Morita}  that the phase is 
stabilized by charge fluctuations enhanced near the Mott transition.

\subsection{spin excitation gap}
\begin{figure}
\includegraphics[width=8.0cm]{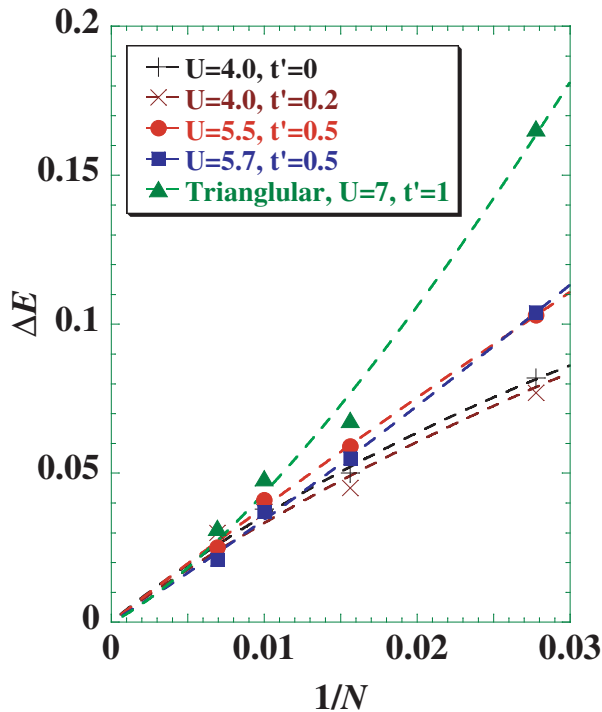}
\caption{(color online) Size scalings of the $S=1$ excitation gaps for several choices of parameters.  The triangles show the case of the model [B] while others are for the model [A].  The dashed curves are fittings to Eq.(6). The circles, squares and triangles are results obtained inside the NMI phase. }
\label{Fig4}
\end{figure}

Now in the NMI phase, system size dependences of the spin excitation gap $\Delta E$ between the ground state and the lowest triplet are shown in Fig.~\ref{Fig4}.  Figure~\ref{Fig4} 
%are rather similar, where the size dependence 
indicates that the triplet excitations become gapless in the thermodynamic limit. 
The gap appears to be scaled asymptotically with the inverse system size $N^{-1}$, namely $\Delta E\sim \zeta/N$.  
At least the extrapolated gap ($< 0.01t$) is much smaller than the typical gap inferred in the spin gapped phase of the corresponding Heisenberg limit ($\ge 0.1J$)~\cite{Sorellaplaquette}. The gapless feature shares some similarity to the behavior in the AFI phase. 
 However, the detailed comparison clarifies a crucial difference as we will show later.  The fitting in the NMI phase to the form (6) gives unphysical values such as $c>1.5$.
We note that the uniform magnetic susceptibility is given by $2/3\zeta$, which implies a nonzero and finite uniform susceptibility.

Except for 1D systems, the present result is the first unbiased numerical evidence for the existence of gapless excitations without apparent long-ranged order in the Mott insulator.
Although a tiny order cannot be excluded if it is beyond our numerical accuracy, in the present NMI phase, the absence of various symmetry breakings including the AF order has already been suggested by the size scalings of the models [A]~\cite{Kashima-Imada} and [B]~\cite{Morita}.  It was shown~\cite{Watanabe} for the model [A] that as well as dimer and  plaquette singlet orders, s- and d-density waves are also numerically shown to be unlikely for four types of correlations probed by
\begin{eqnarray}
C_{\alpha}({\bf q})&=&
\left|
\langle J_{\alpha}({\bf q})J_{\alpha}^{\dagger}({\bf q}) \rangle
\right| 
\nonumber \\
J_{\alpha}({\bf q})&=&\frac{1}{N}\sum_{{\bf k},\sigma}
c^{\dagger}_{{\bf k},\sigma}c_{{\bf k}+{\bf q},\sigma}f_{\alpha}({\bf k})
\label{eq:current}
\end{eqnarray}
with 
\begin{eqnarray}
f_{1}({\bf k})&=&\cos(k_{x})+\cos(k_{y}), \nonumber \\
f_{2}({\bf k})&=&\cos(k_{x})-\cos(k_{y}), \nonumber \\
f_{3}({\bf k})&=&2\cos(k_{x})\cos(k_{y}), \nonumber \\
f_{4}({\bf k})&=&2\sin(k_{x})\sin(k_{y}). 
\end{eqnarray}
%These correlations clarify whether the density wave correlations with four types of symmetry in the form factor grow or not. 
%Figure~\ref{Fig5} in the example of $\alpha=2$ clarifies that the correlation of the d-density wave remains short-ranged. 
%(namely, the staggered flux)~\cite{Affleck} for the model [B].
The calculated results all indicate the absence of long-ranged orders with the correlations staying rather short ranged.
However, it does not strictly exclude an extremely small and nonzero order parameter, although it may melt with further decreasing $U$.

%\begin{figure}
%\includegraphics[width=8.0cm]{ddw.eps}
%\caption{(color online) Size scalings of the staggered flux correlations for the model [B] at $t'/t=1.0$.}
%\label{Fig5}
%\end{figure}

%{\bf Collapse of Dispersions}

\subsection{Lowest $S=1$ and $S=0$ excitations}

%In addition to the above indirect evidence for the appearance of a new phase, 
The lowest energy $S=1$ excitations at each momentum sectors $E_1({\bf k})$ show further dramatic difference between the AFI and NMI phases.  
The lowest energy states with specified momenta, $E({\bf k})$ are calculated from the spin-momentum resolved PIRG. When $E({\bf k})$ becomes the maximum at ${\bf k}_{max}$ and the minimum at ${\bf k}_{min}$, we introduce the width $W\equiv E({\bf k}_{max})-E({\bf k}_{min})$.  

We first make a general remark on $W$ expected from the known phases. Note that in the AFI phases as well as in the stripe phase, the lowest $S=1$ excitations $E_1({\bf k})$ give nothing but the spin wave and the momentum dependence is given by the spin wave dispersion. Therefore, $W$ represents the dispersion width.  On the other hand, in the Fermi liquid, it is given by the lowest edge of the continuum of the Stoner excitations (particle-hole excitations) and in a certain range of the momenta, they are gapless. For example, for the case of noninteracting Fermions at half filling on the square lattice, $W$ becomes zero.  In numerical calculations of the Fermi liquid, however, it has, in general, a finite width $W>0$ due to the finite-size effect and the width is expected to vanish as $W\propto 1/N^{1/d}$, namely being scaled by the inverse linear dimension of the system size.  

In the AFI phase, the dispersion is essentially described by the spin-wave spectrum similar to 
\begin{equation}
\Delta E(k) =4J\sqrt{1-\gamma_k^2}
\end{equation} 
with 
\begin{equation}
\gamma_k=\frac{1}{2}(\cos(k_x)+\cos(k_y))
\end{equation} 
for the spin wave theory of the Heisenberg model, but modified because of finite $U$.  
%The calculated dispersion width ($\sim 1.5$ for $U=4, t'=0$) is comparable to the estimate of the spin-wave dispersion at $J=0.4$.
The calculated dispersion width at $U=4,t=1$, and $t'=0$, shows a small system size dependence and is around 1.5, which can be compared with the dispersion width ($\sim 1.6$ for $J=0.4$) of the ordinary spin wave obtained from the mapping of the Hubbard model to the Heisenberg model~\cite{Coldea}.  

\begin{figure}
\includegraphics[width=8.0cm]{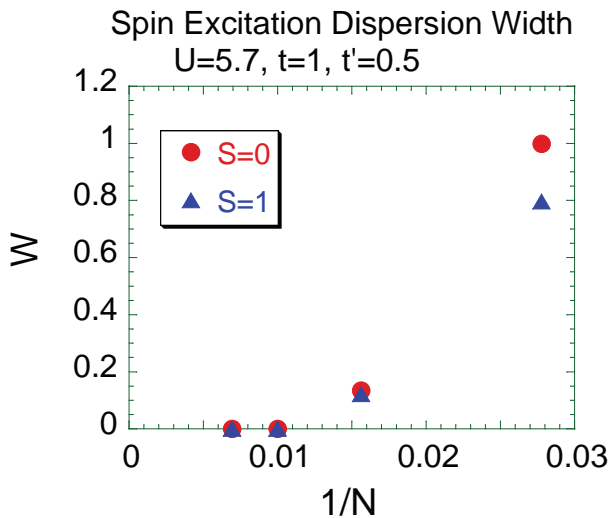}
\caption{(color online) Size scalings of $W$ for $S=0$ and $S=1$ excitations in the NMI phase for the model [A].}
\label{Fig6}
\end{figure}

In marked contrast, the width $W$ for $S=1$ excitation in the NMI phase has strong and monotonic system size dependence as in Fig.~\ref{Fig6}.  For systems larger than $8 \times 8$ lattice, the dispersion becomes vanishingly small.   The vanishing $W$ is consistent with what expected in the Fermi liquid, although the spin liquid phase is certainly insulating. It implies that only the ``spinon" excitations form an excitation continuum in the presence of the charge gap.  Although it is not definitely clear, the size dependence seems to show very quick collapse of the dispersion with increasing system size and may not be fitted by a power of the inverse system size as in the single-particle Stoner excitations in metals.   Such quick collapses of $W$ are observed solely in the NMI phase irrespective of the models (namely commonly seen in the models [A] and [B]).  
The collapse implies that the triplet excitations cannot propagate as a collective mode.  
We also note that the gap of $S=1$ excitations from the ground state, namely $\Delta E$ is scaled by $1/N$ as described above. Therefore, the momentum degeneracy within $S=1$ sectors seems to be much higher than the spin degeneracy in the thermodynamic limit.   

The presence of such degenerate excitations well accounts for the quantum melting of simple translational symmetry breakings including the AF order, because a long-ranged order in the two-dimensional systems is destroyed when the excitation becomes flatter than $\Delta E(k)=k^2$. This is because of the infrared divergence of fluctuations in the form $\int k^{d-1}dk\frac{1}{\Delta E(k)}$ with $d=2$.  

The total singlet state ($S=0$) at any total momentum ${\bf k}$ also shows degenerate structure in the ground state for larger system sizes as in Fig.~\ref{Fig6}. We similarly introduce the width $W$ as $W\equiv E({\bf k}_{max})-E({\bf k}_{min})$ within the singlet $S=0$ sector. The width $W$ vanishes in the NMI phase in the model [A] as well as in [B]. This again implies the excitation continuum, which has larger degeneracy than the spin excitations.

%\subsection{Edward-Anderson's order parameter}
%
%As the index of spin galss, we consider the following Edward-Anderson's order parameter,
%\begin{equation}
%T\left( q \right) = \frac{1}{{3N}}\sum\limits_{i,j}^N {\left| {\left\langle {{\bf{S}}_i  \cdot {\bf{S}}_j } \right\rangle } \right|} e^{iq\left( {R_i  - R_j } \right)} 
%\end{equation}.

\subsection{Spin renormalization factor}

The spin renormalization factor $Z_s(q)$ is defined as 
\begin{equation}
Z_s \left( q \right) = \left| {\left\langle {S = 1,q} \right|Z\left| {S = 0,q = 0} \right\rangle } \right|
\end{equation}
where 
\begin{equation}
Z \equiv \frac{1}{{\sqrt 2 }} \sum\limits_k {\left( 
{c_{k + q, \uparrow }^\dag  c_{k, \uparrow }  - c_{k + q, \downarrow }^\dag  c_{k, \downarrow } 
} \right)}. 
\end{equation}
It is rewritten in the site representation as
\begin{equation}
Z_s \left( q \right) = \frac{1}{{\sqrt 2 }}\left| {\sum\limits_j {\left\langle {S = 1,q} \right|\left( {n_{ \uparrow j}  - n_{ \downarrow j} } \right)\left| {S = 0,q = 0} \right\rangle e^{ - iqR_j } } } \right|.
\end{equation}
The PIRG wave functions for $S=0$ and $S=1$ are given by spin-projection
operator. As the derivation of its matrix element is somewhat lengthy and 
needs spin algebra, it is summarized in Appendix.

\begin{figure}[h]
\includegraphics[width=8.0cm]{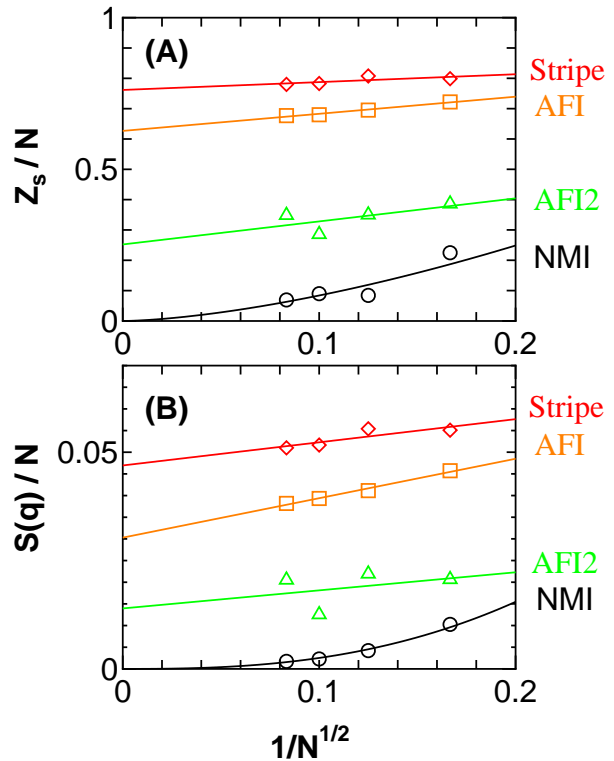}
  \caption{(color online) Size dependence of spin renormalization factor and $S(q)$ at the peak value
  are shown for AFI, new AFI(AFI2), Stripe and NMI phases.}
  \label{sr}
\end{figure}

In analogy with the renormalization factor of the quasiparticle weight in the Fermi liquid, $Z_s$ measures whether the spin excitation is spatially extended and can propagate coherently or not. If $Z_s$ has nonzero values, the spin excitation can propagate coherently. Figure \ref{sr} indeed shows that $Z_s$ remains nonzero for these ordered phases.
In fact, in AFI, AFI2 and stripe phases, the magnons are well-defined elementary excitations in momentum space and spatially propagates coherently, which are reflected in nonzero values of the extrapolated $Z_s$ to the thermodynamic limit. In contrast, the renormalization factor appears to scale to zero for NMI phase, which implies that the spin excitations dressed by other spins are spatially localized.

\section{Discussions and Summary}

The present excitation spectra show the following double-hierarchy structure: Although the singlet ground state is unique, there exist enormous number of low-energy excitations within the $S=0$ sector leading to nearly dispersionless momentum dependence of singlet excitations. Although the energy of lowest-energy state at each total momentum has a substantial momentum dependence in finite-size systems, thus forms a dispersion-like structure arising from the finite-size effects, the momentum dependence quickly collapses with increasing system size.  Another degeneracy near the ground state appears in the excitations with $S>0$. The lowest-energy states with $S=1, S=2, \cdots$ denoted by $E_1, E_2, \cdots$ have a tower-like structure, $E_1<E_2<...$ in finite-size systems. However, these spin excitation gaps again collapse with increasing system size supporting gapless spin excitations in the thermodynamic limit. With increasing system sizes, the singlet excitation energies become vanishing with much faster rate than that of vanishing excitation energies for the excitations to nonzero spins such as triplet. In other words, the collapse of the excitation energies for the spin dependence seems to be much slower than the collapse in the momentum dependence. This generates a hierarchy structure of the excitation energies.  

Here we discuss a possible interpretation of the properties. 
The vanishing gap may allow the following interpretation: The ground state is given by a superposition of dynamical singlet bonds, which cover the whole lattice. Those singlet bonds with vanishingly small singlet binding energy have a nonzero weight, because of the distribution of the singlets over long distance, where, for example, weights decays with a power law with increasing bond distance as in a variational long-ranged RVB wavefunction~\cite{Hsu}. 

  The collapse of $W$ and vanishing renormalization factor $Z_s$ show that the ground state is nearly degenerate with other low-lying states obtained by spatial translations and they have vanishing off-diagonal Hamiltonian-matrix elements each other.  
Although the translational symmetry is retained in the original Hamiltonian, the present orthogonality may imply a large degeneracy near the ground state.
The absence of the spin renormalization factor $Z_s=0$ implies first that the Goldstone mode like magnons in the symmetry broken phase does not exist. Furthermore, it also excludes the existence of coherent single spin excitations as in the idea of the particle-hole-type excitations at the hypothetical spinon Fermi 
surface\cite{Lee,Motrunich}.  The result rather suggests that a spin excitation obtained from the unbinding of the weak singlet may be spatially localized because of the dressing by the surrounding sea of singlets.     

In classical frustrated systems, the spin glass can be stabilized by an infinitesimally small quenched randomness, for example, in the Ising model on a triangular lattice~\cite{Ising}, where the macroscopic degeneracy remains in the regular system. 
In the present case, it may also be true that tiny randomness may further stabilize the freezing of the localization of spins and leads to the spin glass. 

The nature of the gapless and dispersionless excitations is not completely clarified for the moment.  
Although the coherence of the spin excitations must be more carefully examined, the present result supports that an unbound spin triplet does not propagate coherently due to strong scattering by other weakly bound RVB singlets. 
This result also means that the quantum melting of the AF order occurs through the divergence of the magnon (or Goldstone mode) mass.

This NMI phase appears to be stabilized by the Umklapp scattering~\cite{Rice}, where the Mott gap is generated without any symmetry breaking such as antiferromagnetic order. The degenerate excitations within the singlet, which are similar to the present results, but with a spin gap were proposed in the Kagome and pyrochlore lattices based on small cluster studies~\cite{Chalker}. The possible symmetry breaking from degenerate singlets was also examined on the pyrochlore lattice~\cite{Harris,Tsunetsugu}, while the spins were again argued to be gapful in contrast to the present results.  In our results, the gapless spin excitations become clear only at larger system sizes than the tractable sizes of the diagonalization studies.  

We briefly discuss experimental implications of the present new quantum phase.
Recent results by Shimizu {\it et al.}~\cite{Kanoda} on $\kappa$-(ET)$_2$Cu$_2$(CN)$_3$  appear to show an experimental realization of the quantum phase we have discussed in the present work.  In fact this compound can be modeled by a single band Hubbard model on nearly right triangular lattice near the Mott transition.  The NMR relaxation rate  
shows the nonmagnetic and gapless nature retained even at low temperatures ($\sim 0.2$K) and suggests the present quantum phase category. Another organic compound also shows a similar behavior~\cite{Kato}     

Systems with Kagome- (or triangular-) like structures, $^3$He on graphite~\cite{Ishida} and volborthite Cu$_3$V$_2$O$_7$(OH)$_2$$\cdot$2H$_2$O~\cite{Hiroi} show nonmagnetic and gapless behaviors.  
On the other hand, glass-like transitions are seen in 3D systems, typically in pyrochlore compounds as $R_2$Mo$_2$O$_7$ with $R=$Er, Ho, Y, Dy, and Tb~\cite{Taguchi} and in fcc structure, Sr$_2$CaReO$_6$~\cite{Wiebe}.  It is remarkable that the glass behavior appears to occur even when randomness appears to be nominally absent in good quality single crystals of stoichiometric compounds.  
Although the lattice structure, dimensionality and local moments have a diversity, many frustrated magnets show gapless and incoherent (glassy) behavior. The present result on gapless and degenerate structure emerging without quenched randomness offers a consistent concept with this universal trends. It would be an interesting open issue whether the present system leads to such a glass phase at $T=0$ in 2D by introducing randomness.

In summary, we have studied the 2D Hubbard model on two types of square lattices with geometrical frustration effects arising from the next-nearest neighbor transfer $t'$. In the parameter space of the interaction $U$ and $t'$, paramagnetic metal, two antiferromagnetic insulator phases, a stripe-ordered insulator phase, and a new degenerate quantum spin-liquid phase are found.  The quantum spin liquid (in other words, nonmagnetic insulator) phase has gapless spin excitations from the degenerate ground states, and furthermore the dispersionless modes are found in all the spin sectors. The calculated spin renormalization factor suggests that the gapless spin excitation is spatially localized and does not propagate coherently in the thermodynamic limit.
Recent experimental findings, including quantum spin liquids in 2D and spin glasses in 3D suggest a relevance of this phase in disorder-free and frustrated systems. 
\vskip 5mm

\section*{Acknowledgments} 
We would like to thank S. Watanabe for valuable discussions in the early stage of this work and useful comments.
A part of the computation was done at the supercomputer center in ISSP, University of Tokyo.   

\section*{Appendix}
In this appendix, we discuss how to evaluate the matrix elements between 
different spin states.
We consider the state with definite spin $S$ and its $z$-component, which is given from general Slater determinant $\left| \phi \right\rangle$ by spin projection $P_{S,M}$.  The detailed form of $P_{S,M}$ is given by 
$\sum_K g_K L^S_{M,K}$ in Ref.\cite{QPPIRG,Ring}  where $g_K$'s are parameters. 
We denote such spin projected wave function as
$
\left| \phi_{S,M} \right\rangle  \equiv P_{S,M} \left| \phi  \right\rangle 
$.
Now we consider an operator $O^{\lambda}_{\mu}$ which changes spin quantum number 
by $\lambda$ and its $z$-components by $\mu$.
Its expectation value is evaluated by Wigner-Eckart theorem. Therefore we
introduce reduced matrix element 
$
\left\langle {\alpha ,S\left\| {O^\lambda  } \right\|\alpha ',S'} \right\rangle 
$
generally defined by
\begin{widetext}
\begin{equation}
\left\langle {\alpha ,S,M\left| {O_\mu ^\lambda  } \right|\alpha ',S',M'} \right\rangle  = \left(  -  \right)^{S - M} \left( {\begin{array}{*{20}c}
   S & \lambda  & {S'}  \\
   { - M} & \mu  & {M'}  \\
\end{array}} \right)\left\langle {\alpha ,S\left\| {O^\lambda  } \right\|\alpha ',S'} \right\rangle 
\end{equation}
where $\alpha$ and $\alpha'$ are additional quantum numbers.
By considering the commutation relation between spin-projector and the
operator $O^{\lambda}_{\mu}$,
a following formula can be derived \cite{Ring} as
\begin{equation}
\left\langle {\varphi _{S_1 } } \right.\left\| {O^\lambda  } \right\|\left. {\phi _{S_0 } } \right\rangle  = {\textstyle{{\left( {2S_1  + 1} \right)\left( {2S_0  + 1} \right)} \over {8\pi ^2 }}}\sum\limits_{K_0 K_1 \bar K_{1\mu } } {g_{K_0 }^* g_{K_1 } \left(  -  \right)^{S_0  - K_0 } \left( {\begin{array}{*{20}c}
   {S_0 } & \lambda  & {S_1 }  \\
   { - K_0 } & \mu  & {\bar K_1 }  \\
\end{array}} \right)} \int {d\Omega D_{\bar K_1 K_1 }^{S_1 } \left\langle \varphi  \right.\left| {O_\mu ^\lambda  R\left( \Omega  \right)} \right|\left. \phi  \right\rangle } 
\label{transition}
\end{equation}
where, in reduced matrix element, the $z$-component is suppressed,
$D_{M,K }^{S} $ is Wigner's $D$-function and $R\left( \Omega  \right)$
is a rotation operator in spin space \cite{Ring}.

Now we consider the spin renormalization factor. Its operator $O_\mu ^1$
is defined by 
\begin{equation}
O_\mu ^1  = \sum {\left\langle {{\textstyle{1 \over 2}},\sigma ,{\textstyle{1 \over 2}},\sigma '} \right|} \left. {1,\mu } \right\rangle c_\sigma ^\dag  \tilde c_{\sigma '} 
\end{equation}
where $
\tilde c_{j,\sigma }  = ( - )^{{\textstyle{1 \over 2}} - \sigma } c_{j,-\sigma } $. Its three components are 
$O_0^1  = {\textstyle{1 \over {\sqrt 2 }}}\left( { - c_ \uparrow ^\dag  c_ \uparrow   + c_ \downarrow ^\dag  c_ \downarrow  } \right)
$,
$
O_1^1  = c_ \uparrow ^\dag  c_ \downarrow  
$ and
$
O_{ - 1}^1  =  - c_ \downarrow ^\dag  c_ \uparrow  .
$
As the wave function is determined in the half-filled space,
the $z$-component of spin is zero, which means 
$
g_K  = \delta _{K,0}. 
$
Moreover we set $S_0=0$ and $S_1=1$ and normalization factor of projected
wave functions like  
$
\sqrt {{\textstyle{{\left( {2S_0  + 1} \right)} \over 2}}} \sqrt {{\textstyle{{\left( {2S_1  + 1} \right)} \over 2}}}  = \frac{{\sqrt 3 }}{2}
$ 
is also taken into account.
Therefore we can obtain a following
formula as 
\begin{equation}
\left\langle {\varphi _{1,0} } \right.\left| {O_0^1 } \right|\left. {\phi _{0,0} } \right\rangle  =\sum\limits_{\mu  = 0, \pm 1} {\int {\sin \beta d\beta d_{\mu 0}^1 \left( \beta  \right)\left\langle \varphi  \right.\left| {O_\mu ^1 e^{i\beta S_y } } \right|\left. \phi  \right\rangle } } .
\end{equation}
In the nuclear structure physics, the formula Eq.(\ref{transition}) is
often used.
\end{widetext}

%------------------------------------------------------------------------------

\end{document}